# NEW GUIDELINES FOR SPREADSHEETS

*John F. Raffensperger*
*Department of Management, Private Bag 4800*
*University of Canterbury, Christchurch, New Zealand*
*j.raffensperger@mang.canterbury.ac.nz*
*22 May 2001*

**ABSTRACT**

*Current prescriptions for spreadsheet style specify modular separation of data, calculation and output, based on the notion that writing a spreadsheet is like writing a computer program. Instead of a computer programming style, this article examines rules of style for text, graphics, and mathematics. Much "common wisdom" in spreadsheets contradicts rules for these well-developed arts. A case is made here for a new style for spreadsheets that emphasises readability. The new style is described in detail with an example, and contrasted with the programming style.*

## 1. INTRODUCTION

Proper writing of spreadsheets is a special concern to management educators and operations researchers. Educators should teach good habits and must read students' work. Operations research consultants use spreadsheets to produce math models for clients; these models can be large and complicated. Math modelling is now widely taught with spreadsheets, so teachers of math modelling must also teach spreadsheet use (see [Winston, 1996], for example). So we operations researchers have a special interest in good spreadsheet style. While this article has an operations research point of view, it has general application.

Spreadsheets are often hard to read and spreadsheets often contain mistakes. [Panko and Sprague, 1998] found that around 2% of spreadsheet cells contain errors. The likelihood of a correct model less than 50% for even a spreadsheet with only 35 cells. Why are spreadsheets so difficult? Part of the problem is due to existing prescriptions for spreadsheet style.

Most literature about spreadsheet style specifies a spreadsheet style based on computer programming and show a preoccupation with standards. In effect, people who have written about spreadsheet style have in their heads the metaphor that "Writing a spreadsheet is like writing a computer program."

The computer program style is a "standard" format with separate blocks for data, calculation, and output, as in [Bromley, 1985], [Kee, 1988], [Kee and Mason, 1988], [Stone and Black, 1989]. Some articles, such as [Bissell, 1986], [Edge and Wilson, 1990], and [Crain and Fleenor, 1989], mainly recommended that a standard format include various mixes of the following: heading, date, file name display, author, approval signoff line, table of contents, error summary box, instruction area, range names, global protection, absolute references where possible, validation formulas for input and output, no constants in formulas.

Most spreadsheet writers do not have a background in data processing [Cragg and King, 1993], so a "spreadsheet- as-program" metaphor is one that a reader is unlikely to expect, understand, or want. [Cragg and King, 1993] and [Davis, 1996] found that managers like spreadsheets for the freedom from internal information technology groups and may be hostile to perceived interference. So spreadsheet users do not use and do not want this style, and they do not want imposed standards.

Avoiding standards, [Edwards, Finlay, and Wilson, 2000] wrote a good high-level overview of spreadsheet use, guidelines for scoping spreadsheets, and "best practices" for verification. Similar to other writers about spreadsheet style, their article views the spreadsheet as a decision support system. By contrast, this article focuses more on cell-by-cell construction, and for the first time, draws on literature about style for text, mathematics, and graphics.

Abstractly, it is true that a spreadsheet has three parts: data, calculation, and output. However, this functional structure was originally designed for the computer's needs of batch processed punched cards, rather than the reader's needs. The input-calculation-output structure implies the calculation is a black box that the reader should ignore rather than read. Computer code is not written primarily for self-display, but a spreadsheet is meant to be seen and read. Instead of "data and calculation" or even "decision variable and constraint," the reader will be thinking in terms of the business problem, such as "person, shift assignment, preference." This suggests that a spreadsheet should be organised to follow the business logic. To do this, we need a new metaphor to replace "The spreadsheet is a computer program."

[Conway and Ragsdale, 1997] cite and comment on [Kee, 1988], "Reliable spreadsheet software begins with a standard format for developing spreadsheet applications'... However, contrary to Kee's sentiments (given above), for many optimisation problems we find that the forced use of a standard format results in spreadsheets models [sic] that are more difficult to construct, less reliable, and more difficult to understand." Later: "In most cases, we believe the spreadsheet design which communicates its purpose most clearly will also be the most reliable, auditable and modifiable design."

Conway and Ragsdale wrote the *only* paper with ideas other than "have modules for input, computation, and output, and avoid constants in formulas". Their important new proposals are (1) that related formulas should be in physical proximity, and (2) how we should write depends on how we *read.* All previous work found prescribes against such structure or ignores these aspects of layout. This article significantly extends the emphasis on readability by examining how styles for text, graphics, and mathematics can be applied to spreadsheets. The goal is to go beyond personal taste in spreadsheet style, to find well-tested concepts in more developed media, and use those rules for spreadsheets.

A spreadsheet is a mixture of text, graphics and mathematics, a form of expression and computation**.** Rules of style for those forms apply to spreadsheets. For example, as with writing, a spreadsheet is easier to read if its text has proper spelling and grammar. As with graphics, a spreadsheet is easier to read colour is used carefully. And as with mathematics, a spreadsheet should be easier to understand if the formulas are reasonably simplified.

With the gracious permission of Lindo Systems, Inc. (http://www.lindo.com), this paper uses as an example a spreadsheet called Assign.xls. Assign.xls is distributed with *What's Best!,* [Lindo Systems, 19961, a commercial spreadsheet solver. The file can be downloaded with the student version of the company's Solver Suite from http://www.lindo.com. Assign.xls was written by operations researchers who do spreadsheets for a living and have a stake in making them understandable. Assign.xls is a simple employee scheduling problem. It illustrates key concepts of spreadsheet writing for educators and operations researchers.

In the following, a *numeric cell is* a formula or a constant referenced by formulas in other cells. A label, even if it is a numeric constant, is not a numeric cell. Also, we will avoid the word "user". The analogy to writing provides clearer terminology - writer and reader. You will see Excel formulas of the form WB(a, "operator", b), where "operator" is "=", "<=" or ">=". This is how *What's Best!* defines a constraint, where *a* is the left-hand side and *b is* the right-hand side.

**The new guidelines for writing spreadsheets:** Write a spreadsheet as text, mathematics and graphics are written. This paper is organised in order of these recommendations.

- Make your spreadsheets read from left to right and top to bottom.
- Be concise.
- Format for description, not decoration.
- Expose rather than hide information.

## 2. MAKE A SPREADSHEET READ LEFT TO RIGHT AND TOP TO BOTTOM.

2. 1. [Gopen and Swan, 1990] used linguistics and cognitive psychology to study scientific writing. They wrote, "Since we read from left to right, we prefer the context on the left, where it can more effectively familiarize the reader. We prefer the new, important information on the right, since its job is to intrigue the reader. Information is interpreted more easily and more uniformly if it is placed where most readers expect to find it." The contextual information on the left is considered old, and information on the right is considered new. Gopen and Swan are echoed by [Cohen, 1997] and [Microsoft, 1995, p. 384].

How would this apply to a spreadsheet? In a spreadsheet, old information is the input data, since the reader is expected to know it. The new information is the formula, derived information that the reader seeks. We expect to see the data first, and when we have digested that, we expect the output formula, to the right or below nearby. Intermediate formulas logically are data for later formulas, so the rule applies recursively to all numeric cells. Therefore, to write clearly, each formula should depend only on cells above and to the left.

Exceptions depend on reader expectations. An accounting balance sheet typically has years in columns. A year's profit in one column at the bottom may flow to the next year in the next column at the top. The reader of accounting may reasonably expect one year's bottom line to flow into next year.

[Archer, 1989] and [Davis, 1996] observed that cell relationships can be represented as a directed graph, which Archer called a cell relationship diagram. Excel's auditing toolbar has buttons to insert temporary arrows that point to the dependents or precedents of a cell. We will call these graphical segments - Archer's cell relationship diagram - the arcs *of precedence.*

A spreadsheet reads from left to right and top to bottom if, for every numeric cell, all the cell's arcs of precedence start above and to the left of the cell.

Clicking several times on the Trace Precedents button displays the complete precedence tree of the cell. This tree can be quite illuminating. If a precedence tree is tangled like a bowl of spaghetti, the spreadsheet legitimately can be called a spaghetti spreadsheet! If a spreadsheet's precedence tree includes blank cells, the spreadsheet is *perverse,* since it depends on information that is missing.

The Assign.xls Model sheet is shown in Figure 1. The first numeric cell is the Preference Total cell, but no other information about preferences appears on the Model sheet. After hunting, we find the Preferences sheet listed at the bottom. Assign.xls has the objective function at the top and front, the way one might write a Lindo model ([Lindo Systems, 1996]). Preference Total is the objective function, because E4 has range name WBMAX (in the box below the auditing toolbar). The precedence arc is no help; all precedents of this cell are on another sheet. We will come back to Assign.xls.

2.2. Have short arcs of precedence. Old information should be followed immediately by the related new information ([Gopen and Swan, 1990]). [Higham, 1993, p. 15], writing about mathematics, states, "Try to minimize the distance between a definition and its place of first use." [Conway and Ragsdale, 1997] wrote, "Things which are logically related ... should be arranged in close physical proximity and in the same columnar or row orientation."

If a cell is close to its dependents, the reader will more easily see the relationship between them. The spreadsheet will be naturally organised by blocks of meaning, blocks that reflect the business rather than the mechanical requirements of the spreadsheet. By contrast, arranging by input-calculation-output separates related cells by a large visual distance. A cell in an input block and a cell in a calculation block will be far from each other, so their logical relationship is harder to see.

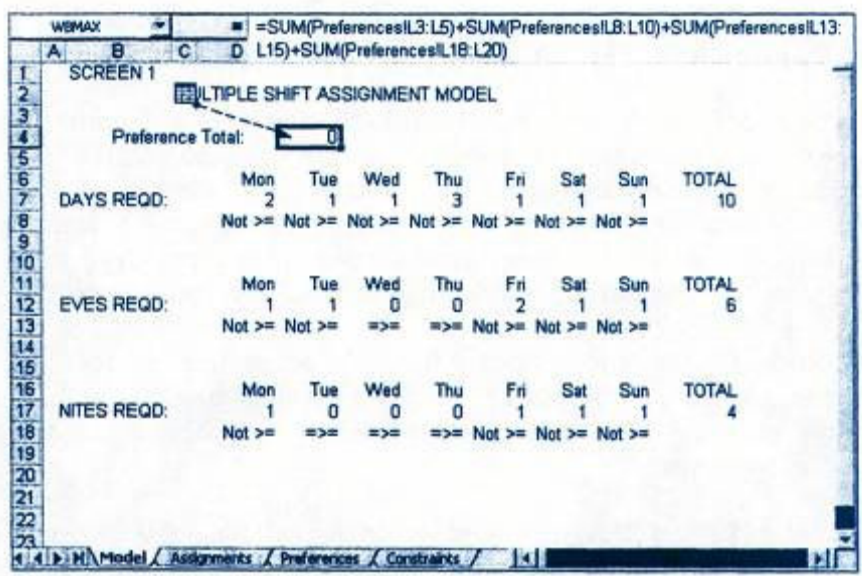

Figure 1. Assign.xls, version 1. The Preference Total precedence arc points meaninglessly off screen. The Preference Total cell is too far from its precedents.

## 3. BE CONCISE.

Regarding text, [Strunk and White, 1979, p. 23] wrote about text, "Vigorous writing is concise." [Higham, 1993, p. 25] wrote about mathematics, "Do not use mathematical symbols unless they serve a purpose." [Tufte, 1983, pp. 51, 100] wrote about graphics, "Graphical excellence is that which gives to the viewer the greatest number of ideas in the shortest time with the least ink in the smallest space..." [Daellenbach, 1994] wrote about math modelling, "A good model is a model that is as parsimonious as possible in terms of the variables/aspects included."

None of the spreadsheet literature was concerned with this; that a spreadsheet should be concise is entirely novel, and controversial. But the literature shows clearly that about 2% of spreadsheet cells are wrong ([Panko and Sprague, 1998]). Reducing the number of cells therefore may increase the probability of a correct spreadsheet.

3. 1. Be concise with sheets. If it will fit on one sheet, consider putting it on one sheet. This is controversial - other writers encourage the use of multiple sheets, with little justification (e.g. [Mather, 1999]). Of course, one sheet may not be optimal for *every* spreadsheet, but multiple sheets are overused. How can multiple sheets cause problems?

Multiple sheets breed unnecessary "spurious" cells and labels, as information is copied from other sheets. A spurious cell has exactly one precedent; it simply points to another cell, such as the formula "B25=A 10" (not in Assign.xls). Mathematically, this is the same as $a = b$. A writer wants to remind the reader of data from another section; later formulas refer to the spurious reference rather than the original data. Such copying can duplicate the entire "input sheet" on every sheet! Now the reader feels the temptation to check back and forth, to verify the formula visually. Various bad habits motivate the use of spurious references, but mainly the problem is from long arcs of precedence or multiple sheets. In text, the item should be edited out as redundant. In graphics, the duplicate visual would be erased. In mathematics, the extra variable should be substituted out algebraically.

Multiple sheets dislocate related blocks; the reader must remember a context from one screen to the next. When moving from one screen to the next, the reader worries, I hope it's laid out the same as the previous one I just spent all this time learning."

Multiple sheets are hard to navigate. Many people do not know the key sequence for changing sheets (Ctrl-Pg Up, Ctrl-Pg Dn) and therefore must use the mouse, but most can use keys to move around a single sheet.

Multiple sheets make information harder to find. The search function in Excel does not scan different sheets by default, but only the selected sheet. To select several sheets at once, the reader must click each sheet tab while pressing Ctrl. Following this, the search will scan the selected sheets. Most people will not know such arcana.

Multiple sheets render auditing tools useless, as we saw above. An arc of precedence to off sheet cells displays an uninformative "off the sheet" icon, so the reader cannot visually see how cells are related. Putting the model on one sheet lets the reader see it in one glance, with the View Zoom command if necessary.

Multiple sheets make it hard to find blanks. Writers frequently do not bother to delete blank sheets in the file (Excel defaults to 16). This can be disconcerting to the reader. Are formulas on Sheet 6? There is no way to know, except to open Sheet 6 and look, a waste of time.

These problems with multiple sheets tend to be overlooked in the name of "modularity." What about large spreadsheets? Isn't it more convenient for large files if they are broken up in different sheets? Of course, many times a modular approach may be called for. However, large spreadsheets can often be reduced in size considerably by moving everything to one sheet, then eliminating the spurious cells.

3.2. Use a minimum of blank space, and only to divide blocks visually. Empty cells are not equivalent to empty space in graphics. Blank space in a picture does not hide anything and is not accidentally covered by nearby information. Its effect is purely visual. By contrast, a spreadsheet cell is a discrete object. It can be formatted as hidden. It can be blank, but perversely another cell can depend upon it. A blank cell may appear to contain data that is really in a different cell. Minimum blank space keeps the spreadsheet on fewer screens, and helps avoid an impression of hidden or misplaced information.

Though empty cells are not equivalent to blank space in a graphic, empty cells can be used in the same way, if the writer is careful. Separate blocks with blanks as one separates sentence's with a period or paragraphs with an indent. But arrange the data types so there is a reasonable minimum of punctuation. Small blocks can be adjacent without blank space between.

3.3. Keep information in one cell logically in one cell visually. Make blank cells look blank. Cells that appear to hold information should hold information. Try to avoid labels that overlap neighbouring cells. Leave on the default grid so cells are displayed.

In Figure 1, the Preference Total label could be in B4, C4, or D4. In fact, both C4 and D4 are blank. The label is in cell B4 with leading spaces. The leading spaces are unnecessary and should be deleted. The label should be put in cell D4, adjacent to the cell it labels.

For Figure 2, cut and paste was used to put Assign.xls on one sheet. Preference Total was moved to the bottom, so its precedents would be above it. Then Excel's default grid lines were turned on, and blank rows and columns were deleted. The spreadsheet now fits in three screens, even at low 640 by 480 resolution.

![Figure 2 spreadsheet screenshot showing Multiple Shift Assignment Model with cell F4 selected containing formula =WB(F12+F16+F20+F24,">=",F3)]

Figure 2 Assign.xls, version 2, on one sheet. Cells in rows 4, 6, and 8 refer to cells below, against the direction we read. B 12, B 16, and B20 also refer to cells below, and are unrelated to cells nearby. The spreadsheet should be arranged so each cell depends only on cells above.

The *What's Best!* constraints cells in row 4 depend on cells below, so the DAYS REQD, EVES REQD, and NITES REQD blocks should be moved down. Similarly, cell B3 is referenced by a cell below and to the right, with no direct bearing on nearby cells. It is like a sentence out of context, or a definition given too early. It should be moved downwards to be near the cells that use it. We will do this for Figure 3.

3.4. Be concise with blocks. Align data types consistently. If "names" are listed column-wise in one place, then list them column-wise everywhere. This gets the reader oriented to viewing the spreadsheet in a consistent way. Break this rule to confuse the reader and waste space with extra labels. Align the primary data type in rows. Excel has 65,536 rows versus 256 columns, so if the primary data type is downwards, the writer is less likely to run out of room. More importantly, we read left to right, so the reader expects to see the table structure across the top, in the column labels. The repetitive information should be below, so the reader views it by paging down. Where possible, have a single table.

[Conway and Ragsdale, 1997] wrote, "A design that results in formulas that can be copied is probably better than one that does not." Four signs for finding such a design are time, label repetition, concreteness, and formula transposition. All numeric cells varying with the same time periods should be in a single table, with the time periods in the far left column (except in accounting balance sheets). Label repetition (e.g. the days of the week labels in Figure 1) suggests that the writer could consolidate blocks. A concrete data type, such as "ex-president", is probably a better structure than an abstract data type, such as "data" or "constraints". Transposing with formulas is a loud sign that data types in the spreadsheet are misaligned. Transposing with formulas makes construction and modification difficult.

When a spreadsheet requires more than one table, stack unrelated blocks vertically (like a web page) or horizontally (like ticker tape), but not both (a bulletin board). Information one screen down and one screen right is hard to find, unless it is part of a table that starts on the far left. A bulletin board layout requires more keystrokes to navigate. Row and column operations will affect other blocks, so changes require moving cells, which is time consuming and error prone. So avoid the bulletin board structure.

Figure 3. Assign.xls version 4, bottom part, with the auditing toolbar. Cells J44, J46, and J48 are dangling, since they have no dependents. These can be deleted.

Assign.xls was originally structured as a linear program, decision variables, constraints, objective coefficients, etc. (A linear program is a set of linear inequalities used to solve a business problem.) The objective function is at the top, as it might be in a linear program. A client is unlikely to want to use it that way. The client instead will probably think about ex presidents. What is Bush's assignment this week? How can a new ex-president be added? Structured by decision variables, preferences, and constraints, the client must copy three sections separately. We will rearrange Assign.xls so the primary data type is an ex-president, and move the "#Work" cell closer to where it is used. If we want to add another ex-president, there is only one block to copy. The amount of visual space used by Assign.xls (as in Figure 3) is now two and a half screens, better than the original four sheets.

## 4. BE CONCISE WITH CELLS.

All cells in a spreadsheet, sometimes even blanks, require the reader's interpretation. Unnecessary cells take up the reader's precious time, time that the writer wants the reader to spend on understanding the business logic. Unnecessary cells can contain mistakes, add visual and logical clutter, add to the bulk of the file, and confuse the reader.

4. 1. Eliminate spurious references (described above). Fortunately, Assign.xls has none.

4.2. Erase dangling cells. A dangling cell is a numeric cell without dependents, a calculation not used anywhere else. A cell not on the precedence tree for the bottom line is dangling. Again, following [Archer's, 1989] cell relationship diagram, dangling cells will be leaves on the graph. Usually, a spreadsheet contains a "bottom line", such as an objective function or balance sheet total profit. Except for this, dangling cells are usually unnecessary.

Just as a spurious reference is analogous to redundant text, a dangling cell is analogous to irrelevant text, an unused variable, or unneeded decoration. Irrelevant verbiage in text is edited out. In graphics, repetitive visuals are erased. In mathematics, unneeded variables are substituted out. So, unless there is a compelling reason to keep them, erase dangling cells.

There are four main types of dangling cells: the unused input, the validation formula, the useless intermediate calculation, and the interpreted output. An unused input is a relic from an earlier version or an error in the model. (Whether a relic or an error, the spreadsheet needs more work.) We shall see more about relics later.

Authors cited earlier recommended formulas to check the validity of input and output. These extra formulas are dangling by definition. Temporary validation formulas are useful for debugging, but reduce readability, and eventually should be erased. Writers insert useless intermediate calculations, thinking the reader "will want to see a subtotal, just for their information". But the writer should instead drive home the main point rather than display trivial side matters. Let the readers put these in as they wish.

Interpreted output is like a comedian repeating the punch line, hoping for an extra laugh. For example, consider the formulas (not in Assign.xls):
      D49: SUM(D3:D48)
      D50: if(D49>0, "Surplus of " +TEXT(D49,0), "0")
      D51: if(D49<0, "Funding gap of "+TEXT(-D49,0), "0")

Cells D50 and D51 merely interpret D49. Cell D49 should be labelled like "Surplus (gap)", and the two spurious formulas D50 and D51 should be erased. Avoid labels containing formulas. While they seem clever to the writer, the reader has a harder time distinguishing the numeric model from the documentation. Use formulas only for the model and use concise constant text for documentation.

Does Assign.xls have dangling cells? In the NIXON block of Figure 3, cell range C40:H42, and cells I40, J35 and J36 do not appear on the precedence tree of the objective function Total Preferences. But they are required for the constraints, and are referenced by the *What'sBest!* solver. This is a compelling reason to leave them in. But the "Total" cells J44, J46, and J48 are danglers of the useless-intermediate-calculation variety. They have no dependents and they are not constraints for the model, so they will be erased.

4.3. Simplify formulas. For example, the formula (not from Assign.xls) C6*(A4) + A6*C6 + ((C6*A5)) could be simplified to C6*(A4+A5+A6), or C6*SUM(A4:A6), which changes automatically if a row were inserted. Table 1. summarises issues of formula readability.

| More readable | Less readable |
|---|---|
| Similar terms repeat and reference similar ranges. | Irregular terms reference various ranges, references appear more than once. |
| References read left to right in row and column order. | References are ordered randomly in the formula. |
| Well-known formulas (e.g. the quadratic formula) are in a single cell. | Well-known formulas are made cryptic by separation into multiple cells. |
| Division generally appears at the end. | Division appears randomly. |
| Referenced cells contain constants. | Referenced cells contain formulas or blanks. |
| The formula has the fewest characters necessary. | The formula has unnecessary parentheses and spaces; the formula can be simplified. |

Table 1 Factors in cell readability.

Simplifying a formula requires knowledge of the order of precedence for arithmetic operators. The formula (C7/A8)*A7 could be simplified to C7/A8*A7 because multiplication and division are commutative. However, the reader may wonder if the writer meant CV(A8*A7), which is not the same. So put division last: C7*AVA8.

It is a good habit to use parentheses when we are not sure of the precedence of the arithmetic operators. However, in a long formula, extraneous characters can accumulate to make the formula hard to understand. When teaching students how to write a spreadsheet, educators ought to take a few minutes to remind students about the rules of precedence for arithmetic operators. Use the fewest characters necessary to write the formula correctly.

4.4. Nest and erase formulas where appropriate. Sometimes making a few cells slightly more complicated allows many cells to be erased. Nesting is especially appropriate when a formula in cell x has only one dependent, cell y. The formula in x can be copied and substituted for the address of cell x in cell y. This process should stop when the formula in the dependent cell y begins to lose readability.

Other authors (e.g. [Freeman, 1996] and [Mather, 19991]) say the opposite: complicated formulas should be separated into multiple cells. However, nesting of formulas follows from mathematical writing, which prescribes substituting out unnecessary variables. Virtually *every* other form of expression favours brevity. Why do we consider spreadsheet verbosity a virtue? The writer should weigh the number of cells against the readability of individual formulas, but the error thus far has been on too many cells.

Many cells can be eliminated with the sumproduct () formula. In Assign.xls, Figure, cells J37, J38, and J39 of the Nixon block multiply preferences by assignments, and then are summed in Preference Total. J37, J38, and J39 could be replaced with one sumproduct() and then nested directly into Preference Total. We can do the same substitution with the other blocks. The Preference Total becomes
C51: =SUMPRODUCT(C4:I6,C7:I9) + SUMPRODUCT(C14:I16,C17:I19) + SUMPRODUCT(C24:I26,C27:I29) + SUMPRODUCT(C34:I36,C37:I39)

This allows us to erase twelve cells, 4% of the numeric cells. All precedents of this cell are now constants, and may be viewed in a one click. Hence, it will be easier to debug and edit. While long, this formula is readable because it is regular.

4.5. Eliminate relics from earlier versions. A relic is anything no longer needed, but appears because the writer has not cleaned it up. Relics confuse the reader.

A simple way to look for spreadsheet relies is to press the End key, then the Home key. This moves the cursor to the lowest and furthest right cell for which computer memory has been allocated. If this last cell is not the bottom right cell of the intended spreadsheet, then there may be relics. Most of the time, extra rows and columns contain old formats.

A fast way to see if apparently blank rows or columns affect the spreadsheet is to delete them. If #REF! errors appear, undo the change and find the problem; formulas perversely refer to blank cells. References to blank are as dangerous as vagueness in legal writing.

Assign.xls contains relics of earlier versions. In the Model sheet, the data seem to be in columns A through J. But width was adjusted of columns as far right as AU. In fact, in the original version, pressing End and Home moves the cursor to cell IT22, far away from the apparent last cell in L 18. Fortunately, these relics are just old formats, as we see in Figure 4.

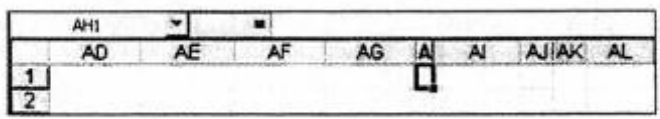

Figure 4. Assign.xls contains relics, unneeded formats from earlier versions.

## 5. FORMAT FOR DESCRIPTION, NOT DECORATION.

A format is descriptive if it displays information and if the reader already knows the format. A format is decorative if it displays no information. Writers format cells for various reasons: to get attention, to try to look professional, from perceived necessity, to encode data types, and to make a cell self-descriptive. These overlap somewhat, and are variously decorative or descriptive. The worst are cryptic; the best are crucial. In the following, we applied to spreadsheets Tufte's suggestions about graphics ([Tufte, 1983, p.183]).

Use formats to get attention or to try to look professional very sparingly. When a writer decorates with formatting, the writer should follow rules of style for those elements. Most spreadsheet writers do not have a degree in graphic arts, and should not attempt to decorate a spreadsheet. For the writer, default formats take no time to apply. For the reader, the default format is best in most cases.

Use only one font size. Large fonts take up screen space, take time to produce, and require changes in row heights which affect the entire spreadsheet width. Avoid using many colours;

readers bring their own complex interpretations and perceptions of colours, and may be colour blind. Avoid multiple styles in one cell, e.g. bold and italic, and use these rarely. Underscored spreadsheet text is ugly. To look professional, strive for a clean look rather than distracting big titles and heavy colours. To get attention on a few specific cells, use a logical layout and concise labels ([Tufte. 1983, p. 183]).

Use formats of perceived necessity very sparingly. The most common format of necessity is the column width. Except for the A column, which should contain labels, try to give all columns about the same width. Many different column widths can be disconcerting; the reader may try to find meaning in a cell width when there is none. Uniform columns are more aesthetically pleasing. Avoid varying row heights. A thick row will tend to force the reader to read primarily downwards. For the same reason, do not rotate fonts ([Tufte, 1983, p. 183]). Do not shrink fonts either - the reader's display may be small.

Format constant data differently from formulas. The most crucial structural distinction is the one between constants and formulas ([Conway and Ragsdale, 1997]). Constants can be interpreted faster than formulas, so the reader will see obvious constants with relief. The old approach to making this distinction was to put input data in a separate block, but this produces long arcs of precedence.

Without a compelling reason, do not try to encode data types beyond the distinction between constants and formulas. Additional encoding will not naturally convey information and can easily become cryptic ([Tufte, 1983, p. 183]). Instead, use a good block layout and clear labelling to make other distinctions between data types.

Following the principle of minimal formatting, either constants or formulas may be formatted - whichever results in the least formatting - with a light grey background colour and a thin grey border. A grey background is easy to notice, photocopies better than a coloured font, and is not distracting. In any case, document the difference on the spreadsheet with a formatted label, e.g. "Formulas are grey." In Figure 5, the formulas have been formatted.

Figure 5 Assign.xls, final version. Abbreviations were eliminated. Formulas are grey.

[Edwards and Finlay, 1997] also distinguish between constant data and formulas by protecting formulas to prevent accidentally overwriting a formula. Their argument is legitimate, but formula protection can be irritating, especially if a formula is wrong and password protected. Cell protection is a hidden format, so other formatting is required to indicate which cells may be changed. Readers should be informed that the spreadsheet is protected and told how to turn it off.

Which cells are constant in Assign.xls? The constant cells are the Assignments and Preferences blocks for each ex-president (Figure). With proper formatting, the reader would probably see that Assign.xls is rather simple. Most cells are preference inputs or decisions.

Use formats of self-description liberally. Format numbers according to their meaning, using widely understood conventions to make information self-descriptive. Right justify numeric cells and the labels above them. All cells of a given data type should display the same number of decimal places - the reader wants a number to self- descriptively indicate its magnitude by its width. If the number is a percent, format it as a percent, so it displays "%". If the number is money, format it to display a monetary symbol, such as "$". Show cents only if they are significant, otherwise decrease the number of displayed decimal places to zero. Show one decimal place ("$24.4") if the number is not in ones (e.g., if it is thousands), and label the cell, row, or column appropriately (e.g., with "(000)"). Show separators ("1,000,000").

## 6. EXPOSE RATHER THAN HIDE INFORMATION.

Put labels in the spreadsheet, and make sure most of them are on the left. On mathematical writing, [Higham, 1993, p. 25] recommends, "Avoid starting a sentence with a mathematical expression, particularly if the previous sentence ended with one, otherwise the reader may have difficulty parsing the sentence." The reader expects the label on the left, not a formula.

Spell out labels and use the spell checker. Abbreviations inhibit comprehension ([Tufte, 1983, p. 183]). Even supposedly "commonly understood" symbols (such as $Q$ for the order quantity, $A$ for order cost) should be written out in a spreadsheet (Order quantity Q, Order cost A).

Use proper case. Text is *not* clearer or easier to read if written in capitals ([Tufte, 1983, p. 183]). Just the opposite - it is harder to find the beginning and end of a sentence, and it is not what readers expect. Capitals are wider than lower case, so less text can fit in the width of a cell. The last version of Assign.xls appears in Figure 5, with proper case.

Of course, there is the old adage, "Do not put constants in a formula," since data in a formula hides information rather than exposes it, as in Table 2. But there is a more powerful rule: try to make every formula reference only constants.

|  Unfriendly           |  Friendly     |
|-----------------------|---------------|
| B5: 100000            | B5: 7%        |
| B6: 20                | B6: 20        |
| B7: PMT(O.07, B6, B5) | B7: $100,000  |
| B8: PMT(B5, B6, B7)   |               |

Table 2 On the left, 0.07 should be in its own cell; references to B5, B6, and B7 should be in order. Cells should have self-descriptive formats.

A spreadsheet with hidden cells is perverse, because hidden cells are inaccessible dependents. Hiding cells or preventing changes in a spreadsheet is irritating and tends to reduce the reader's confidence in the spreadsheet. If a reader takes the time to audit a spreadsheet with hidden cells, the model cannot be proved correct, because the reader cannot see the formulas. The writer assumes the reader cannot improve the work, and preventing derivations from it diminishes its utility. The logic is literally hidden. Hiding cells is like writing in invisible ink.

Password protect can be over-used. Writers password-protect a spreadsheet to hide the entire spreadsheet from unauthorised readers, to hide portions of the spreadsheet from unauthorised readers, to prevent a reader from changing part or all of the spreadsheet (especially formulas), or to track changes to a spreadsheet. Excel allows the writer to password protect the spreadsheet in such a way that the writer can track changes made by others in a shared spreadsheet. Tracking changes is good. Hiding information tends to be bad.

Some writers feel the need to "idiot proof" a spreadsheet. This usually involves lots of formatting, heavy lines around the inputs, many colours, an input sheet, a summary sheet, password protection, etc. Instead, have the fewest cells necessary to produce the result, flow the logic from top to bottom and left to right, and put related cells close together. And put it all on one sheet. "Keep it simple" could be restated as "Keep it small."

Take the time to get it right. Text, graphics, and mathematics require editing. A large spreadsheet may need to be rewritten several times, as we find that we have misoriented data types, understood the problem better, or found a more succinct way to express the model. Make your spreadsheet resemble a system that your reader already knows. If the spreadsheet is automating or extending an existing system, the reader may be familiar with a related form ([Stewart and Flanagan, 1987]).

## 7. CONCLUSION

While we have used analogies to writing text, mathematics, and graphics, the analogy to writing text is perhaps the best. Creating a spreadsheet is more like writing text than it is like coding in C. We should tell students not to embed numeric constants in formulas, but we might explain it by saying "because you should not hide a key definition in a footnote or appendix." Hopefully, this paper has helped to debunk the idea that a "spreadsheet as computer program" structure will improve spreadsheet readability and reduce spreadsheet error. In its place, we proposed a new style for writing spreadsheets based on writing text, graphics, and math models. The style emphasises conciseness and readability.

## 8. ACKNOWLEDGEMENT

Thanks to Linus Schrage and the gentle folk at Lindo Systems for their help.